\documentclass[final,5p,twocolumn]{elsarticle}

\usepackage{amsmath,amssymb}
\usepackage{bm}
\usepackage{graphicx}
\usepackage{booktabs}
\usepackage[colorlinks=true,allcolors=blue]{hyperref}

\journal{Journal of Subatomic Particles and Cosmology}
\begin{document}

\begin{frontmatter}

\title{Hadron masses and pseudoscalar meson form factors in a Poincar\'e-invariant constituent quark model}

\author{Jongheon Baek}
\ead{bjh6941@yonsei.ac.kr}
\address{Department of Physics and Institute of Physics and Applied Physics, Yonsei University, Seoul 03722, Korea}

\begin{abstract}
We study a Poincar\'e-invariant constituent quark model with pair, fine-structure, and finite-range three-quark interactions treated in the diagonalization.
A fit of 20 parameters to 36 meson and 27 baryon masses gives an rms deviation of 6.85~MeV, with the common offset fixed by a spectral condition.
The same mass-fit wave functions give the elastic form factors of ten pseudoscalar mesons in the light-front impulse approximation with the Melosh rotation.
The pion and charged-kaon radii are 0.407 and 0.411~fm, below the measured values. Unlike experiment, the kaon radius slightly exceeds the pion radius.
Replacing the pion radial function by the normalized $S$-wave component of the $\rho$ raises its radius to 0.604~fm, showing that hyperfine compression accounts for most of the deficit in the mean-square charge radius.
With the light-quark size fixed by the pion, the $K^+$ and $K^0$ radii require strange-quark sizes that differ by 2.2 standard deviations.
\end{abstract}


\end{frontmatter}

\section{Introduction}
\label{sec:intro}

The constituent quark model with pair interactions motivated by one-gluon exchange (OGE) has been used for a long time to describe the hadron spectrum~\cite{Godfrey1985,CapstickIsgur:1986,CapstickRoberts2000PPNP}.
In this kind of model, the masses of many mesons and baryons can be reproduced with a modest number of parameters.
Relativistic kinematics, fine structure, and three-quark forces allow a detailed fit of the spectrum.
The transfer of the resulting wave functions to observables outside that fit is a separate test of the mass operator and current.

In Refs.~\cite{Baek2026PPNP,Baek2026JSPC}, we studied the finite-range three-quark interaction in the $\mathrm{S^2AJ}$ model, which has OGE-type color-Coulomb and color-spin interactions and linear confinement~\cite{park2019,noh2021}.
In the present work, we consider the fixed-particle-number Poincar\'e-invariant extension of this model.
The rest-frame mass operator contains the relativistic kinetic energy, the spin--orbit and tensor interactions, and the three-quark interaction, and all of them are treated in the diagonalization.
We fit 20 parameters to 36 meson and 27 baryon masses, with the common offset $D$ fixed by the spectral condition in Sec.~\ref{sec:offset}.
We then calculate the elastic form factors and charge radii of pseudoscalar mesons in the light-front impulse approximation~\cite{ChungCoesterPolyzou1988,Jaus1991,ChoiJi1999,Hwang2002Radii}.
The sensitivity of the pion form factor to the high-momentum wave function and to constituent-quark form factors was studied in Refs.~\cite{Cardarelli1995PLB,Cardarelli1996PRD}.
Here we test the mass-fit wave functions using the pion radius together with both kaon charge radii.
Once the light-quark size is fixed by $\pi^+$, the $K^+$ and $K^0$ radii give two determinations of the strange-quark size.

\section{Model}
\label{sec:model}

\subsection{Intrinsic mass operator}
The intrinsic momenta satisfy $\sum_i\mathbf p_i=0$.
Following Ref.~\cite{Baek2026PPNP}, we write the mass operator as
\begin{align}
 \mathcal M&=\sum_i\sqrt{\mathbf p_i^2+m_i^2}+L^{3Q}_{\mathrm Y,\mathrm{ext}}\nonumber\\
 &\quad+\sum_{i<j}\big(V^C_{ij}+V^{CS}_{ij}+V^{LS}_{ij}+V^{ALS}_{ij}+V^T_{ij}\big).
 \label{eq:pi_mass}
\end{align}
The three-quark term $L^{3Q}_{\mathrm Y,\mathrm{ext}}$ is present only for baryons, and the rest masses are included in the square roots.
Following Ref.~\cite{Baek2026PPNP}, we use $\mathcal J$ for the total angular momentum of a hadron and $J$ for the total orbital angular momentum, so that $\mathcal J$ is obtained by coupling $J$ and the total spin $S$.

\subsection{Front-form Bakamjian--Thomas construction}
\label{sec:bt}
Equation~\eqref{eq:pi_mass} is embedded in the front-form Bakamjian--Thomas (BT) representation of the Poincar\'e group~\cite{BakamjianThomas1953,KeisterPolyzou1991,Coester1992PPNP}.
With $P^\pm=P^0\pm P^3$, seven generators, $P^+$, $\mathbf P_\perp$, $J^3$, $K^3$, and the two transverse light-front boosts, leave the null plane $x^+=0$ invariant and are kinematic, while the light-front Hamiltonian $P^-$ and the two transverse rotations are dynamical~\cite{Dirac1949,Hiller2016PPNP}.
For free particles, all ten generators can be written as functions of $P^+$, $\mathbf P_\perp$ and their conjugate positions, of the free mass $\mathcal M_0=\sum_i\sqrt{\mathbf p_i^2+m_i^2}$, and of the front-form spin $\mathbf j_f$ of the system.
The Poincar\'e commutation relations then follow only from the canonical commutators of these momenta and positions, from the SU(2) algebra of $\mathbf j_f$, and from the fact that $\mathbf j_f$ and $\mathcal M_0$ commute with the momenta and positions and with each other~\cite{KeisterPolyzou1991}.
In the BT construction, $\mathcal M_0$ is replaced by $\mathcal M=\mathcal M_0+V$ in the same functions.
The commutation relations are kept, and a unitary representation of the Poincar\'e group is obtained, if
(i) $V$ commutes with $P^+$, $\mathbf P_\perp$, and their conjugate positions, that is, $V$ acts only on the internal variables,
(ii) $V$ commutes with $\mathbf j_f$, and
(iii) $\mathcal M$ is self-adjoint and positive~\cite{KeisterPolyzou1991,Polyzou2010BT}.
The kinematic generators retain their free form, while the transverse rotations and
\begin{equation}
 P^-=\frac{\mathcal M^2+\mathbf P_\perp^2}{P^+}
 \label{eq:pi_pminus}
\end{equation}
contain $\mathcal M$.
The interaction enters through the mass operator, with $P^+P^--\mathbf P_\perp^2=\mathcal M^2$.

We write $V$ in terms of the canonical rest-frame variables, that is, the momenta $\mathbf p_i$ with $\sum_i\mathbf p_i=0$, the relative coordinates, and the canonical spins.
The light-front internal variables, the momentum fractions and the relative transverse momenta, and the light-front spins of the constituents are obtained from them by a change of variables and by Melosh rotations~\cite{Melosh1974}.
This unitary map, which we call $U$, depends only on the internal momenta, preserves $\mathcal M_0$, and takes the canonical internal angular momentum $\boldsymbol{\mathcal J}_0$ (the sum of the orbital angular momenta and the spins in the rest frame) to $\mathbf j_f$~\cite{ChungCoesterPolyzou1988,PolyzouGlockleWitala2013}.
The conditions (i)--(iii) for $U\mathcal MU^\dagger$ are equivalent to the following conditions for $\mathcal M$: $V$ depends only on the internal variables, $[\mathcal M,\boldsymbol{\mathcal J}_0]=0$, and $\mathcal M$ is self-adjoint and positive.
All terms of Eq.~\eqref{eq:pi_mass}, which are given in Secs.~\ref{sec:pair} and \ref{sec:threeq}, are built from $r_{ij}$, $\mathbf p_i$, and the scalar products $\boldsymbol\sigma_i\cdot\boldsymbol\sigma_j$, $\mathbf L_{ij}\cdot(\mathbf S_i\pm\mathbf S_j)$, and $S_{12}$, so the first two conditions are satisfied.
The matrices of $\mathcal M$ are Hermitian, and all eigenvalues in this work are positive; the lowest corresponds to the pion.
If the complete magnetic multiplets are kept, we obtain a unitary representation of the Poincar\'e group in the retained space.
For three particles, we use the BT construction without the further unitary transformations needed for strong cluster separability~\cite{KeisterPolyzou2012Cluster}.

Because $\mathcal M$ commutes with $P^+$ and $\mathbf P_\perp$, light-front eigenstates can be constructed directly from the mass eigenstates.
If $\psi_{n\mathcal J\mu}$ is an eigenfunction of $\mathcal M$ in the rest-frame variables, $\mathcal M\psi_{n\mathcal J\mu}=M_n\psi_{n\mathcal J\mu}$, then
\begin{equation}
 |\Psi_{n\mathcal J\mu};P^+,\mathbf P_\perp\rangle=|P^+,\mathbf P_\perp\rangle\otimes U|\psi_{n\mathcal J\mu}\rangle
 \label{eq:pi_lf_state}
\end{equation}
is an eigenstate of $P^-$ with the eigenvalue $(M_n^2+\mathbf P_\perp^2)/P^+$, and $\mu$ is the light-front spin projection.
The dynamical eigenvalue problem is the diagonalization of $\mathcal M$ in Sec.~\ref{sec:fit}, and its eigenvalues $M_n$ are the hadron masses.
The kinematic map $U$ gives the light-front internal wave function $U\psi_{n\mathcal J\mu}$ independently of the external momenta $P^+$ and $\mathbf P_\perp$.
This internal wave function is the same in the different forms of dynamics~\cite{Polyzou2010BT}, and the explicit form of $U$ for the mesons is given in Sec.~\ref{sec:em}.

\subsection{Pair interaction}
\label{sec:pair}
We define $m_{ij}=2m_im_j/(m_i+m_j)$, twice the reduced mass. The inverse smearing length $\mu_{ij}$ is given by
\begin{equation}
\begin{aligned}
 \mu_{ij}&=m_0\exp\!\big[\alpha_{\rm sm}+\beta_{\rm sm}(m_{ij}/m_0)^{3/10}\big],\\
 \rho_\mu(r)&=\frac{\mu^3}{\pi^{3/2}}e^{-\mu^2r^2},\qquad m_0=1~\mathrm{GeV}.
\end{aligned}
 \label{eq:pi_smearing}
\end{equation}
Here $F_i^c=\lambda_i^c/2$ $(c=1,\dots,8)$ are the color matrices of quark $i$, and $\bar F^c=-F^{cT}$ for an antiquark.
In the color-singlet states, $-\frac34F_i^cF_j^c$ is $1$ for $q\bar q$ and $1/2$ for a quark pair in a baryon.
We also define
\begin{equation}
\begin{aligned}
 K^V_{ij}&=e^{\xi\chi_{ij}},\qquad \chi_{ij}=\Big(\frac{m_i-m_j}{m_i+m_j}\Big)^2,\\
 K^{CS}_{ij}&=(m_{ij}/m_0)^{3/10},\\
 \mu_{2,ij}&=(\mu_{ij}^{-2}+a_2^{-2})^{-1/2},\\
 \mu_{h,ij}&=(\mu_{ij}^{-2}+a_{\rm hf}^{-2})^{-1/2}.
\end{aligned}
 \label{eq:pi_ranges}
\end{equation}
The central and color-spin interactions are
\begin{align}
 V^C_{ij}(r)&=-\frac34F_i^cF_j^c\Big[-K^V_{ij}\,\kappa\,\frac{\mathrm{erf}(\mu_{ij}r)+\mathrm{erf}(\mu_{2,ij}r)}{r}\nonumber\\
 &\qquad+\sigma r-D+v_Ge^{-r^2/r_G^2}+v_He^{-r^2/r_H^2}\Big],
 \label{eq:pi_central}\\
 V^{CS}_{ij}(r)&=-\frac34F_i^cF_j^c\,\frac{K^{CS}_{ij}}{m_im_j}
 \big[g_1\rho_{\mu_{ij}}(r)+g_2\rho_{\mu_{h,ij}}(r)\big]\,\boldsymbol\sigma_i\cdot\boldsymbol\sigma_j .
 \label{eq:pi_contact}
\end{align}
The Coulomb term has one strength $\kappa$ with the two smearing ranges $\mu_{ij}$ and $\mu_{2,ij}$.
The short Gaussian range is fixed at $r_G=3/(20~\mathrm{GeV})=0.02960~\mathrm{fm}$.
These pair functions are common to the meson and baryon sectors.

For the fine structure, we use
\begin{equation}
 \widetilde m_i=\sqrt{m_i^2+\Lambda_F^2},\qquad
 U_{1,ij}(r)=\!\int\! d^3r'\,\rho_{\mu_{ij}}(|\mathbf r-\mathbf r'|)\,\sigma r',
 \label{eq:pi_fine_regulator}
\end{equation}
and one vector-like radial kernel,
\begin{equation}
\begin{aligned}
 \mathcal V_{ij}(r)&=-K^V_{ij}\,\kappa\,\frac{\mathrm{erf}(\mu_{ij}r)+\mathrm{erf}(\mu_{2,ij}r)}{r}+U_{1,ij}(r),\\
 W_{ij}(r)&=\mathcal V'_{ij}(r)/r.
\end{aligned}
 \label{eq:pi_fine_kernel}
\end{equation}
The spin--orbit, antisymmetric spin--orbit, and tensor operators are
\begin{align}
 V^{LS}_{ij}&=-\frac34F_i^cF_j^c\,c_{LS}\Big[\frac{\widetilde m_i^{-2}+\widetilde m_j^{-2}}{4}+\frac{1}{\widetilde m_i\widetilde m_j}\Big]\nonumber\\
 &\qquad\times W_{ij}\,\mathbf L_{ij}\cdot(\mathbf S_i+\mathbf S_j),\nonumber\\
 V^{ALS}_{ij}&=-\frac34F_i^cF_j^c\,\frac{c_{LS}}{4}\big(\widetilde m_i^{-2}-\widetilde m_j^{-2}\big)W_{ij}\,\mathbf L_{ij}\cdot(\mathbf S_i-\mathbf S_j),\nonumber\\
 V^{T}_{ij}&=-\frac34F_i^cF_j^c\,\frac{c_{LS}}{12\,\widetilde m_i\widetilde m_j}\Big(\frac{\mathcal V'_{ij}}{r}-\mathcal V''_{ij}\Big)S_{12},
 \label{eq:pi_fine}
\end{align}
where $\mathbf L_{ij}=\mathbf r_{ij}\times(m_j\mathbf p_i-m_i\mathbf p_j)/(m_i+m_j)$.
All three terms come from the same kernel $\mathcal V_{ij}$, and we use one strength $c_{LS}$ for them.
Because $V^{LS}$ and $V^{ALS}$ have the same coefficient $c_{LS}$, the coefficient of $\mathbf L_{qQ}\cdot\mathbf S_Q$ is proportional to $c_{LS}W_{qQ}\big[1/(2\widetilde m_Q^2)+1/(\widetilde m_q\widetilde m_Q)\big]$, and it vanishes in the heavy-quark limit~\cite{Neubert1994,Korner1994PPNP}.
The scale $\Lambda_F$ limits the inverse-mass enhancement of the light quarks, while $\widetilde m_Q/m_Q\to1$ for the heavy quarks.
Equation~\eqref{eq:pi_fine} is an effective prescription; QCD potential matching imposes additional relations between the central and spin-dependent potentials~\cite{Gromes1984}.

\subsection{Three-quark interaction}
\label{sec:threeq}
For baryons, the finite-range three-quark term is based on $L^{3Q}_{\mathrm Y}$ of Ref.~\cite{Baek2026PPNP}, with modified mass weights:
\begin{align}
 L^{3Q}_{\mathrm Y,\mathrm{ext}}&=f_{\mathrm Y}\big(A_RQ_A^R+CQ_C\big),\nonumber\\
 f_{\mathrm Y}&=\exp\Big[-\sum_{i<j}m_{ij}r_{ij}\Big],\nonumber\\
 Q_A^R&=\sum_{\mathrm{cyc}(l,i,j)}\Big(\frac43\lambda_i^c\lambda_j^c+2\,d^{abc}\lambda_1^a\lambda_2^b\lambda_3^c\Big)\big(w_l-m_0^{-1}\big),\nonumber\\
 Q_C&=\sum_{\mathrm{cyc}(l,i,j)}\Big(\frac43\lambda_i^c\lambda_j^c+2\,d^{abc}\lambda_1^a\lambda_2^b\lambda_3^c\Big)\nonumber\\
 &\qquad\times w_l\Big(\frac{\boldsymbol\sigma_l\cdot\boldsymbol\sigma_i}{m_lm_i}+\frac{\boldsymbol\sigma_l\cdot\boldsymbol\sigma_j}{m_lm_j}\Big),
 \label{eq:pi_threebody}
\end{align}
with the weight
\begin{equation}
 w_i=\frac{m_{i0}}{m_0^2},\qquad m_{i0}=\frac{2m_im_0}{m_i+m_0}.
 \label{eq:pi_weight}
\end{equation}
Here $\lambda_i^c=2F_i^c$, $d^{abc}$ is the symmetric SU(3) tensor, and $(l,i,j)$ runs over the cyclic permutations of $(1,2,3)$.
$Q_A^R$ and $Q_C$ are the color--color and color--spin operators $L^{C-C}$ and $L^{C-S}$ of Ref.~\cite{Baek2026PPNP} for a baryon, where the inverse mass $1/M_l$ is replaced by $w_l-m_0^{-1}$ and $w_l$, respectively.
In the color-singlet baryon, $\langle\lambda_i^c\lambda_j^c\rangle=-8/3$ and $\langle d^{abc}\lambda_1^a\lambda_2^b\lambda_3^c\rangle=80/9$~\cite{Baek2026PPNP}, so the color operator in Eq.~\eqref{eq:pi_threebody} is $128/9$ for every $l$, and
\begin{equation}
\begin{aligned}
 Q_A^R&=\frac{128}{9}\sum_i\big(w_i-m_0^{-1}\big),\\
 Q_C&=\frac{128}{9}\sum_{i<j}\frac{w_i+w_j}{m_im_j}\,\boldsymbol\sigma_i\cdot\boldsymbol\sigma_j .
\end{aligned}
 \label{eq:pi_threebody_singlet}
\end{equation}
$f_{\mathrm Y}$ is the Yukawa-type profile of Ref.~\cite{Baek2026PPNP}, and $m_{i0}$ is defined analogously to $m_{ij}$.
The weight satisfies $w(m_0)=m_0^{-1}$ and approaches $2/m_0$ as $m_Q\to\infty$.
This gives a finite spin-independent contribution to the light cloud of the singly heavy baryons, while the spin terms with the heavy quark keep the inverse-mass suppression.
With the fitted masses, $w_q=0.45$, $w_c=1.28$, and $w_b=1.68~\mathrm{GeV}^{-1}$.

\subsection{Spectral condition for \texorpdfstring{$D$}{D}}
\label{sec:offset}
Because $D$ is multiplied by $-\frac34F_i^cF_j^c$ in Eq.~\eqref{eq:pi_central}, it shifts all meson masses by $-D$ and all baryon masses by $-3D/2$, so a single parameter $D$ sets both offsets.
We fix $D$ by the spectral condition
\begin{equation}
\begin{aligned}
 D&=\min_{a\in\mathcal G}M_a(D=0)-1~\mathrm{GeV},\\
 \mathcal G&=\{q\bar q\,{}^3P_0,\ q\bar s\,{}^3P_0,\ q\bar q\,{}^1P_1,\ (q\bar s)_{1^+,\mathrm{lower}}\}.
\end{aligned}
 \label{eq:pi_floor}
\end{equation}
The scale $E_*=1$~GeV places the lowest of these four levels at $E_*$ and sets the relative meson--baryon offset.
In the fit, the lowest of the four levels is $q\bar q\,{}^3P_0$. Table~\ref{tab:lightp} compares the remaining light $P$ levels with experiment. The $b_1$, $a_1$, and $a_2$ masses, outside the fit, are lower by 216, 97, and 88~MeV. The corresponding $K_1(1270)$, $K_1(1400)$, and $K_2^*(1430)$ masses are lower by 42, 134, and 67~MeV. The 1~GeV condition sets a scalar mass scale rather than fitting this multiplet, and it leaves substantial errors in the light $P$ spectrum.

The long-distance light $q\bar q$ Coulomb coefficient is $2\kappa=2.385$.
For an unsmeared two-body kinetic operator $2\sqrt{\mathbf p^2+m_q^2}$, the Coulomb critical coupling is $4/\pi$~\cite{Herbst1977,Hall2001Salpeter}.
In the present model the Coulomb potential instead tends to the finite value $-2\kappa(\mu_{ij}+\mu_{2,ij})/\sqrt\pi$ at the origin, so the smeared Coulomb attraction is bounded below even though $2\kappa>4/\pi$.

\begin{table}[t]
\centering
\caption{Light $P$-wave masses outside the fit, in MeV. Experimental values are from the 2026 PDG: $b_1(1235)$, $a_1(1260)$ (mass estimate), and $a_2(1320)$ for $q\bar q$; $K_1(1270)$, $K_1(1400)$, and $K_2^*(1430)^0$ for $q\bar s$~\cite{PDG2026}. Scalar assignments are left open.}
\label{tab:lightp}
\footnotesize
\setlength{\tabcolsep}{2.3pt}
\begin{tabular}{lrr@{\hspace{3mm}}lrr}
\toprule
$q\bar q$ & $M$ & Expt. & $q\bar s$ & $M$ & Expt. \\
\midrule
${}^3P_0$ & 1000.0 & -- & ${}^3P_0$ & 1127.2 & -- \\
${}^1P_1$ & 1013.5 & $1229.5(32)$ & $1^+_{\rm lower}$ & 1213.7 & $1256(6)$ \\
${}^3P_1$ & 1132.8 & $1230(40)$ & $1^+_{\rm upper}$ & 1269.5 & $1403(7)$ \\
${}^3P_2$ & 1230.0 & $1318.2(6)$ & ${}^3P_2$ & 1365.3 & $1432.4(13)$ \\
\bottomrule
\end{tabular}
\end{table}

\section{Fit}
\label{sec:fit}

The equal-weight fit includes all 63 masses in Table~\ref{tab:masses}: 36 mesons, 24 ground-state baryons, and three baryon excitations.
There are 20 fitted parameters in addition to the offset $D$ determined by Eq.~\eqref{eq:pi_floor}.
The baryon excitations are assigned to the lowest $D$-dominant $\mathcal J^P=5/2^+$ levels for $\Lambda_c(2880)$ and $\Lambda_b(6152)$ and the $3/2^+$ level for $\Lambda_b(6146)$.
The fit also includes $\eta_c(2S)$, $\psi(2S)$, the $\chi_{c\mathcal J}$ multiplet, and the narrow open-charm $P$-wave states.

The meson space contains $S$, $P$, and $D$ waves, including the ${}^3S_1$--${}^3D_1$ tensor mixing.
The meson wave functions are expanded with up to 56 Gaussians whose ranges are between 0.005 and 40~GeV$^{-1}$, and the kinetic energy is calculated in momentum space.
For baryons, we use the Gaussian expansion method~\cite{hiyamaGEM}.
The orbital sectors are $(SS)_{J=0}$, $(PP)_{J=0}$, $(DS)_{J=2}$, $(SD)_{J=2}$, and $(PP)_{J=2}$ in the three Jacobi rearrangements, and they are coupled to the allowed spin--flavor states at fixed $\mathcal J^P$~\cite{Baek2026PPNP}.
All terms of Eq.~\eqref{eq:pi_mass}, including $L^{3Q}_{\mathrm Y,\mathrm{ext}}$, are included in the diagonalization.
The least-squares fit recalculates $D$ from Eq.~\eqref{eq:pi_floor} at each parameter point.
We minimize the sum of squared mass deviations using numerical derivatives.
The inverse-curvature correlation matrix gives $\rho(a_{\rm hf},g_2)=-0.994$, $\rho(v_H,\kappa)=+0.988$, $\rho(c_{LS},\kappa)=-0.967$.
These correlations show that the masses determine combinations of the couplings more accurately than the individual couplings.

Experimental masses follow Ref.~\cite{Baek2026PPNP}, with $M_{\Xi_b(5955)^-}=5955.5$~MeV from the 2026 PDG~\cite{PDG2026}.

The 20 parameters are listed in Table~\ref{tab:params}.
The rms deviation is 6.85~MeV (mesons 7.63, baryons 5.65~MeV).
Enlarging the meson basis from 40 to 56 Gaussians and the baryon expansions from at most 10 to 12 ranges per Jacobi coordinate changes every mass by at most 0.6~MeV.
The largest residuals in Table~\ref{tab:masses} are $D_{s1}(2536)$ ($-19.3$~MeV), $\psi(2S)$ ($+17.5$~MeV), $D_1(2420)$ ($-15.9$~MeV), and $\Lambda_c(2880)$ ($+11.9$~MeV).

The positive Gaussian term forms a repulsive core of range $r_G=0.02960$~fm and suppresses the wave function at small separations. With $v_G=26.2$~GeV, the light $q\bar q$ central potential tends to 25.4~GeV at the origin. Removing the core while keeping the other parameters and $D$ fixed lowers the pion mass from 141 to 88~MeV and produces a collapsed spin-singlet $b\bar b$ state at 0.49~GeV (rms separation 0.022~fm), driven by short-range color-spin attraction reinforced by the Coulomb attraction. The fitted central potential has its minimum near 0.21~fm. Its local slopes at $r=0.5$, 1.0, and 1.5~fm are 1.65, 0.88, and 0.64~GeV/fm. The asymptotic tension $\sigma=0.085$~GeV$^2$ is about half the $0.18$~GeV$^2$ used in conventional linear potentials~\cite{ChoiJi1999}. The smeared Coulomb and Gaussian terms contribute to the slope at hadronic distances.

\begin{table}[t]
\centering
\caption{The 20 fitted parameters and common offset $D$.
$r_G=3/(20~\mathrm{GeV})$ and $m_0=1$~GeV are fixed.
$^\ast$\,$D$ is fixed by the spectral condition in Eq.~\eqref{eq:pi_floor}.}
\label{tab:params}
\footnotesize
\setlength{\tabcolsep}{4pt}
\begin{tabular}{lr@{\hspace{7mm}}lr}
\toprule
$m_q$ (GeV) & $0.28834$ & $\alpha_{\rm sm}$ & $-1.7871$ \\
$m_s$ (GeV) & $0.51551$ & $\beta_{\rm sm}$ & $2.6924$ \\
$m_c$ (GeV) & $1.7813$ & $g_1$ & $0.57867$ \\
$m_b$ (GeV) & $5.1603$ & $g_2$ & $3.1114$ \\
$\kappa$ & $1.1926$ & $a_{\rm hf}$ (GeV) & $0.28114$ \\
$\sigma$ (GeV$^2$) & $0.08527$ & $c_{LS}$ & $0.22107$ \\
$a_2$ (GeV) & $0.36508$ & $\Lambda_F$ (GeV) & $0.55180$ \\
$\xi$ & $-0.06848$ & $A_R$ (GeV$^2$) & $0.03412$ \\
$v_G$ (GeV) & $26.160$ & $C$ (GeV$^4$) & $-0.000469$ \\
$v_H$ (GeV) & $0.98832$ & $D$ (GeV) & $-0.18244$$^\ast$ \\
$r_H$ (GeV$^{-1}$) & $0.97491$ & & \\
\bottomrule
\end{tabular}
\end{table}
\begin{table*}[t]
\centering
\caption{Experimental masses $M^{\rm exp}$ and values $M$ calculated with the larger basis, in MeV, with $\Delta M=M-M^{\rm exp}$.
The experimental charge states include
$\pi^\pm$, $\rho^\pm$, $K^0$, $K^{*0}$, $D^0$, $D^{*\pm}$, $B^0$,
$D_1(2420)^0$, $D_2^*(2460)^0$, $B_2^*(5747)^0$, and $\Xi_b(5955)^-$.}
\label{tab:masses}
\footnotesize
\setlength{\tabcolsep}{3.2pt}
\begin{tabular}{lrrr@{\hspace{5mm}}lrrr@{\hspace{5mm}}lrrr}
\toprule
State & $M^{\rm exp}$ & $M$ & $\Delta M$ & State & $M^{\rm exp}$ & $M$ & $\Delta M$ & State & $M^{\rm exp}$ & $M$ & $\Delta M$ \\
\midrule
\multicolumn{4}{c}{\emph{Mesons}} & $\chi_{c2}(1P)$ & 3556.17 & 3549.11 & $-7.06$ & \multicolumn{4}{c}{\emph{Baryons (cont.)}} \\
$\pi$ & 139.57 & 141.23 & $+1.66$ & $h_c(1P)$ & 3525.37 & 3513.74 & $-11.63$ & $\Xi^*$ & 1535.00 & 1531.14 & $-3.86$ \\
$\rho$ & 775.11 & 774.01 & $-1.10$ & $\Upsilon(2S)$ & 10023.40 & 10032.64 & $+9.24$ & $\Omega$ & 1672.45 & 1664.50 & $-7.95$ \\
$K$ & 497.61 & 491.76 & $-5.85$ & $\chi_{b0}(1P)$ & 9859.44 & 9859.42 & $-0.02$ & $\Lambda_c$ & 2286.46 & 2282.47 & $-3.99$ \\
$K^*$ & 895.81 & 900.66 & $+4.85$ & $\chi_{b1}(1P)$ & 9892.78 & 9891.69 & $-1.09$ & $\Sigma_c$ & 2453.97 & 2452.15 & $-1.82$ \\
$\phi$ & 1019.46 & 1016.51 & $-2.95$ & $\chi_{b2}(1P)$ & 9912.21 & 9910.44 & $-1.77$ & $\Sigma_c^*$ & 2518.48 & 2522.68 & $+4.20$ \\
$D$ & 1864.84 & 1867.89 & $+3.05$ & $h_b(1P)$ & 9899.30 & 9897.63 & $-1.67$ & $\Omega_c$ & 2695.20 & 2701.85 & $+6.65$ \\
$D^*$ & 2010.26 & 2013.34 & $+3.08$ & $\Upsilon_2(1D)$ & 10163.70 & 10170.33 & $+6.63$ & $\Omega_c^*$ & 2765.90 & 2765.03 & $-0.87$ \\
$D_s$ & 1968.30 & 1972.52 & $+4.22$ & $\psi(2S)$ & 3686.10 & 3703.56 & $+17.46$ & $\Xi_{cc}$ & 3621.20 & 3625.47 & $+4.27$ \\
$D_s^*$ & 2112.10 & 2108.81 & $-3.29$ & $D_1(2420)$ & 2420.80 & 2404.92 & $-15.88$ & $\Lambda_b$ & 5619.60 & 5608.74 & $-10.86$ \\
$\eta_c$ & 2984.10 & 2987.24 & $+3.14$ & $D_2^*(2460)$ & 2461.10 & 2456.29 & $-4.81$ & $\Sigma_b$ & 5815.64 & 5813.45 & $-2.19$ \\
$J/\psi$ & 3096.90 & 3096.76 & $-0.14$ & $D_{s1}(2536)$ & 2535.11 & 2515.78 & $-19.33$ & $\Sigma_b^*$ & 5834.74 & 5841.21 & $+6.47$ \\
$B$ & 5279.72 & 5268.05 & $-11.67$ & $D_{s2}^*(2573)$ & 2569.10 & 2563.05 & $-6.05$ & $\Omega_b$ & 6046.10 & 6042.90 & $-3.20$ \\
$B^*$ & 5325.20 & 5325.28 & $+0.08$ & $B_2^*(5747)$ & 5739.60 & 5748.92 & $+9.32$ & $\Xi_c$ & 2470.50 & 2482.23 & $+11.73$ \\
$B_s$ & 5366.77 & 5358.40 & $-8.37$ & $B_{s2}^*(5840)$ & 5839.88 & 5849.68 & $+9.80$ & $\Xi_c'$ & 2578.30 & 2582.26 & $+3.96$ \\
$B_s^*$ & 5415.40 & 5416.19 & $+0.79$ & \multicolumn{4}{c}{\emph{Baryons}} & $\Xi_c^*$ & 2646.00 & 2649.56 & $+3.56$ \\
$B_c$ & 6274.47 & 6284.71 & $+10.24$ & $N$ & 938.27 & 936.74 & $-1.53$ & $\Xi_b$ & 5797.00 & 5800.52 & $+3.52$ \\
$\eta_b$ & 9398.70 & 9396.17 & $-2.53$ & $\Lambda$ & 1115.68 & 1118.54 & $+2.86$ & $\Xi_b'$ & 5934.90 & 5932.85 & $-2.05$ \\
$\Upsilon(1S)$ & 9460.40 & 9450.94 & $-9.46$ & $\Sigma$ & 1197.45 & 1194.58 & $-2.87$ & $\Xi_b^{*-}$ & 5955.50 & 5960.50 & $+5.00$ \\
$\eta_c(2S)$ & 3637.80 & 3631.90 & $-5.90$ & $\Delta$ & 1232.00 & 1227.96 & $-4.04$ & $\Lambda_c(2880)$ & 2881.63 & 2893.55 & $+11.92$ \\
$\chi_{c0}(1P)$ & 3415.50 & 3415.98 & $+0.48$ & $\Xi$ & 1321.71 & 1330.72 & $+9.01$ & $\Lambda_b(6152)$ & 6152.50 & 6149.40 & $-3.10$ \\
$\chi_{c1}(1P)$ & 3510.67 & 3505.28 & $-5.39$ & $\Sigma^*$ & 1383.70 & 1385.37 & $+1.67$ & $\Lambda_b(6146)$ & 6146.20 & 6140.66 & $-5.54$ \\
\bottomrule
\end{tabular}
\end{table*}

\begin{figure*}[!t]
\centering
\includegraphics[width=0.94\textwidth]{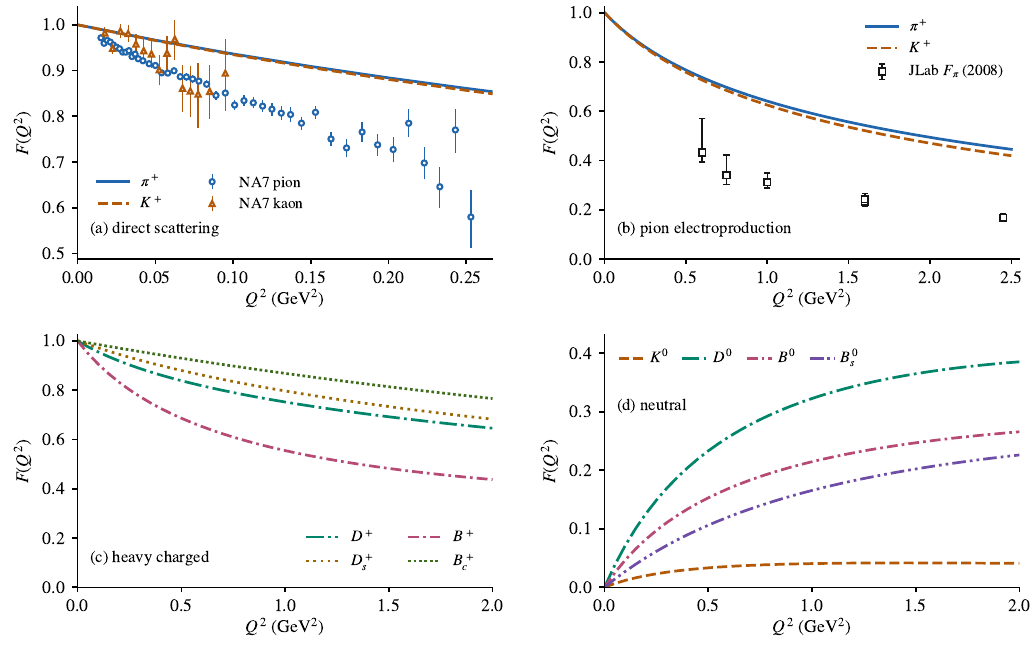}
\caption{Elastic form factors from the fit to the 63 masses: (a) light charged mesons at low $Q^2$, (b) light charged mesons over the JLab range, (c) heavy charged mesons, and (d) neutral mesons.
The points in (a) are the 45 pion bins of NA7~\cite{Amendolia1986Pion} and the 15 charged-kaon bins of Ref.~\cite{Amendolia1986Kaon}, as tabulated in HEPData.
The NA7 values are shown as the positive square root of $|F|^2$, with $F(0)=1$.
Their statistical errors in $|F|^2$ are transformed to errors in $F$; the common normalization uncertainties of 0.9\% (pion) and 1\% (kaon) in $|F|^2$ are kept separate and omitted from the bars.
The six JLab pion points in (b) are from Ref.~\cite{Huber2008Fpi}; the experimental and asymmetric extraction-model uncertainties are added in quadrature.}
\label{fig:ff}
\end{figure*}

\section{Pseudoscalar meson form factors}
\label{sec:em}

We use the same pseudoscalar eigenfunctions and calculate the elastic form factors with pointlike Dirac constituent currents in the $q^+=0$ light-front valence approximation~\cite{ChungCoesterPolyzou1988,Jaus1991,ChoiJi1999,Hwang2002Radii}.
For a meson, the map $U$ of Sec.~\ref{sec:bt} is written explicitly as follows.
The rest-frame relative momentum $\mathbf k$ is expressed with the light-front variables $(x,\mathbf k_\perp)$ as
\begin{align}
 M_0^2&=\frac{m_1^2+k_\perp^2}{x}+\frac{m_2^2+k_\perp^2}{1-x},\nonumber\\
 k_z&=\Big(x-\frac{1}{2}\Big)M_0+\frac{m_2^2-m_1^2}{2M_0},\nonumber\\
 \phi(x,\mathbf k_\perp)&=\sqrt{\frac{\partial k_z}{\partial x}}\,\psi(\mathbf k),\qquad
 \frac{\partial k_z}{\partial x}=\frac{E_1E_2}{x(1-x)M_0},
 \label{eq:pi_lf_wavefunction}
\end{align}
where $x$ is the light-front momentum fraction of constituent 1 and $E_i=\sqrt{m_i^2+\mathbf k^2}$.
Here $\psi$ is the ${}^1S_0$ radial eigenfunction of $\mathcal M$ times $Y_{00}$, and we normalize $\psi$ with $d^3k$ and $\phi$ with $dx\,d^2k_\perp$.
The constituent spins are rotated by Melosh rotations~\cite{Melosh1974,Jaus1991,ChoiJi1999}
\begin{equation}
 R_M(x,\mathbf k_\perp;m)=\frac{m+xM_0-i\boldsymbol\sigma\cdot(\hat{\mathbf z}\times\mathbf k_\perp)}{\sqrt{(m+xM_0)^2+k_\perp^2}},
 \label{eq:pi_melosh}
\end{equation}
where we use $(x,\mathbf k_\perp;m_1)$ for constituent 1 and $(1-x,-\mathbf k_\perp;m_2)$ for constituent 2.
Then the light-front wave function $U\psi$ of the pseudoscalar meson is
\begin{align}
 \Phi_{\lambda_1\lambda_2}(x,\mathbf k_\perp)&=\phi(x,\mathbf k_\perp)\sum_{s_1s_2}
 \langle\tfrac12s_1\tfrac12s_2|00\rangle\nonumber\\
 &\quad\times\langle\lambda_1|R_M^\dagger(x,\mathbf k_\perp;m_1)|s_1\rangle\nonumber\\
 &\quad\times\langle\lambda_2|R_M^\dagger(1-x,-\mathbf k_\perp;m_2)|s_2\rangle,
 \label{eq:pi_lf_spin}
\end{align}
where $\lambda_i$ are the light-front spin projections.
Equations~\eqref{eq:pi_lf_wavefunction}--\eqref{eq:pi_lf_spin} implement the kinematic map of the mass eigenfunction $\psi$.
The plus component of a pointlike Dirac current preserves the light-front spin of the struck constituent, so the form factor is the overlap of the initial and final $\Phi$.
With $\mathcal A(x)=(1-x)m_1+xm_2$ and the charges $e_i$ in units of the positron charge, the sum over $\lambda_1$ and $\lambda_2$ gives
\begin{align}
 F(Q^2)=\sum_{i=1}^2e_i&\int_0^1dx\int d^2k_\perp\,
 \phi^*(x,\mathbf k'_{\perp i})\phi(x,\mathbf k_\perp)\nonumber\\
 &\times\frac{\mathcal A^2+\mathbf k_\perp\cdot\mathbf k'_{\perp i}}
 {\sqrt{(\mathcal A^2+k_\perp^2)(\mathcal A^2+k_{\perp i}'{}^2)}},\nonumber\\
 \mathbf k'_{\perp1}=\mathbf k_\perp+(1-x)&\mathbf q_\perp,\quad
 \mathbf k'_{\perp2}=\mathbf k_\perp-x\mathbf q_\perp,\quad Q^2=\mathbf q_\perp^2 .
 \label{eq:pi_form_factor}
\end{align}
The mean-square charge radius is $\langle r^2\rangle=-6(\hbar c)^2F'(0)$, and we use it also for the neutral mesons.
In the numerical calculation, Eq.~\eqref{eq:pi_form_factor} is evaluated in the rest-frame variables of the unprimed wave function, and $|F(0)-e_1-e_2|<10^{-8}$.

\begin{table*}[t]
\centering
\caption{Mean-square charge radii and elastic form factors of pseudoscalar mesons from the
light-front impulse current, Eq.~\eqref{eq:pi_form_factor}, with the wave functions from the mass fit.
The row from Ref.~\cite{Hwang2002Radii} is its finite-quark-mass light-front result (Table 2), in fm$^2$.
The experimental radii are the squares of the PDG charge radii of $\pi^\pm$ and $K^\pm$ and the PDG
mean-square radius of $K^0$~\cite{PDG2026}.}
\label{tab:em}
\footnotesize
\setlength{\tabcolsep}{3.4pt}
\begin{tabular}{lrrrrrrrrrr}
\toprule
 & $\pi^+$ & $K^+$ & $K^0$ & $D^+$ & $D^0$ & $D_s^+$ & $B^+$ & $B^0$ & $B_s^0$ & $B_c^+$ \\
\midrule
$\langle r^2\rangle$ (fm$^2$) & $+0.1660$ & $+0.1693$ & $-0.0333$ & $+0.1187$ & $-0.1938$ & $+0.0700$ & $+0.2571$ & $-0.1267$ & $-0.0681$ & $+0.0351$ \\
$F(0.5~\mathrm{GeV}^2)$ & $0.7672$ & $0.7583$ & $0.0332$ & $0.8385$ & $0.2332$ & $0.8800$ & $0.6863$ & $0.1529$ & $0.1061$ & $0.9298$ \\
$F(1~\mathrm{GeV}^2)$ & $0.6411$ & $0.6251$ & $0.0405$ & $0.7520$ & $0.3228$ & $0.7969$ & $0.5549$ & $0.2147$ & $0.1656$ & $0.8682$ \\
$\langle r^2\rangle$ Ref.~\cite{Hwang2002Radii} & -- & -- & -- & $+0.1840$ & $-0.3040$ & $+0.1240$ & $+0.3780$ & $-0.1870$ & $-0.1190$ & $+0.0433$ \\
$\langle r^2\rangle_{\rm exp}$ (fm$^2$) & $+0.434(5)$ & $+0.314(35)$ & $-0.077(10)$ & -- & -- & -- & -- & -- & -- & -- \\
\bottomrule
\end{tabular}
\end{table*}

The results are shown in Table~\ref{tab:em} and Fig.~\ref{fig:ff}, including the direct comparison with NA7 pion and kaon data~\cite{Amendolia1986Pion,Amendolia1986Kaon} and the JLab pion results~\cite{Huber2008Fpi}.
The unit-charge radius contributions in the $D$ mesons are 0.0218~fm$^2$ for the charm quark and 0.313~fm$^2$ for the light antiquark. In $B^+$ the corresponding light-quark and $\bar b$ values are 0.384 and 0.0037~fm$^2$.

The finite-mass light-front calculation of Ref.~\cite{Hwang2002Radii} gives $\langle r^2\rangle_{D^+}=0.184$ and $\langle r^2\rangle_{B^+}=0.378$~fm$^2$, compared with 0.1187 and 0.2571~fm$^2$ here. The signs agree, while the six heavy--light mean-square-radius magnitudes are 32--44\% smaller. 

The charged pion and kaon radii are 0.407 and 0.411~fm, compared with $0.659(4)$ and $0.560(31)$~fm; the $K^0$ mean-square radius is $-0.0333$~fm$^2$, compared with $-0.077(10)$~fm$^2$~\cite{PDG2026}. Unlike experiment, the kaon radius slightly exceeds the pion radius.
The pion mass-fit wave function has $\langle\mathbf k^2\rangle^{1/2}=0.96$~GeV and rest-frame rms separation 0.39~fm; for the kaon these values are 0.87~GeV and 0.42~fm. Section~\ref{sec:limit} compares these distributions and the effect of current dressing.

\section{Limitations for the light mesons}
\label{sec:limit}

First, the impulse calculation uses pointlike constituent currents. Quark form factors, anomalous magnetic moments, two-body currents, and meson-cloud contributions can modify the charge radii~\cite{Wang2023PPNP,RamalhoPena2024PPNP}.
In the front form, the interaction enters $P^-$ and the transverse rotation generators, and the one-body impulse result depends on the frame and current component~\cite{ChungCoesterPolyzou1988,KeisterPolyzou1991}.
For spin-zero states, one can define the composite electromagnetic current $J^\mu$~\cite{FangKubisKupsc2021PPNP} by
\begin{equation}
 \langle P'|J^\mu(0)|P\rangle=(P'+P)^\mu F(Q^2),\qquad Q^2=-(P'-P)^2.
 \label{eq:pi_composite_current}
\end{equation}
This defines a covariant, conserved elastic matrix element using the form factor extracted in the $q^+=0$ frame, but it does not determine the interaction-dependent current.

Second, the photon couples to a quark through the dressed quark--photon vertex.
In the Dyson--Schwinger approach, the Ball--Chiu vertex, which is constructed from the dressed quark propagator, gives less than half of $r_\pi^2$, and the remainder comes from the $\rho$ pole of the dressed vertex~\cite{MarisTandy2000vertex}.
In light-front models, this dressing has been included by the constituent-quark form factors~\cite{Cardarelli1995PLB} or by the vector-meson dominance of the vertex~\cite{deMelo2006}.
To estimate its size, we multiply the current of constituent $i$ in Eq.~\eqref{eq:pi_form_factor} by the monopole form factor $f_i(Q^2)=(1+Q^2/m_{V_i}^2)^{-1}$, with $m_{V_i}=m_\rho$ for the $u$ and $d$ quarks and $m_{V_i}=m_\phi$ for the $s$ quark~\cite{PDG2026}.
Then the radius becomes
\begin{equation}
 \langle r^2\rangle=\sum_ie_i\big(\langle r^2\rangle_i+s_i\big),\qquad s_i=\frac{6(\hbar c)^2}{m_{V_i}^2},
 \label{eq:pi_quark_size}
\end{equation}
where $\langle r^2\rangle_i$ is the mean-square radius from pointlike constituent $i$, as in Sec.~\ref{sec:em}, and $s_{u,d}=0.389$ and $s_s=0.225$~fm$^2$.
These monopoles give $\langle r^2\rangle=0.555$, $0.503$, and $-0.088$~fm$^2$ for $\pi^+$, $K^+$, and $K^0$, compared with $0.434(5)$, $0.314(35)$, and $-0.077(10)$~fm$^2$. 
For the hadronic pion form factor, Klingl, Kaiser, and Weise found that pion loops and a momentum-dependent $\gamma\rho$ coupling raise the pion mean-square radius from the standard VMD value of $0.40$ to $0.44$~fm$^2$, in agreement with experiment~\cite{Klingl1996}.

The pion and both kaon radii test one additive size per constituent flavor. Fixing the light-quark size by the pion gives $s_{u,d}=0.268$~fm$^2$. The charged and neutral kaons then require $s_s=-0.10\pm0.10$ and $+0.14\pm0.03$~fm$^2$. The difference is 2.2 standard deviations, including the shared pion-radius covariance and treating the three measured radii as independent.

Light-front calculations of the measured pion radius also provide a wave-function comparison~\cite{Jaus1991,ChoiJi1999}.
For example, in Ref.~\cite{ChoiJi1999}, the Gaussian parameter is determined without the hyperfine interaction, and the hyperfine interaction is added as a perturbation.
The $\pi$ and $\rho$ then have the same wave function, with $\langle\mathbf k^2\rangle^{1/2}=\sqrt{3/2}\,\beta_{q\bar q}=0.39\text{--}0.45$~GeV; the present $\rho$ value is 0.43~GeV.
With this wave function, the measured $\pi^+$, $K^+$, and $K^0$ radii are reproduced reasonably well with the pointlike constituents~\cite{ChoiJi1999}.
On the other hand, the wave functions of a mass operator with the one-gluon-exchange interaction have a high-momentum tail, which strongly affects the pion form factor~\cite{Cardarelli1995PLB,Cardarelli1996PRD}.
The spin--spin term carries the factor $1/(m_im_j)$, so its heavy-quark contribution decreases with the heavy mass. The pseudoscalar/vector rest-frame rms-separation ratios are 0.53 for $\pi/\rho$, 0.65 for $K/K^*$, 0.84 for $D/D^*$, and 0.93 for $B/B^*$. For light quarks, the attractive hyperfine interaction compresses the pion to about half the rms separation of the $\rho$, whereas Ref.~\cite{ChoiJi1999} uses the same radial function for both. Replacing the pion radial function in Eq.~\eqref{eq:pi_form_factor} by the normalized $S$-wave component of the $\rho$, while retaining the pseudoscalar spin factor and fitted quark mass, raises the charge radius from 0.407 to 0.604~fm.
After this radial substitution, only 0.070~fm$^2$ remains missing; adding the full monopole contribution of 0.389~fm$^2$ would overshoot the measured pion mean-square charge radius.

In the mass operator, the small pion mass should reflect its Goldstone-boson nature under dynamical chiral symmetry breaking~\cite{CloetRoberts2014PPNP,Roberts2021PPNP,Giacosa2025PPNP,Swanson2012PPNP}.
This is the case in the Nambu--Jona-Lasinio model~\cite{NJL1961,Klevansky1992,HatsudaKunihiro1994,VoglWeise1991PPNP} and in the Dyson--Schwinger and Bethe--Salpeter approaches~\cite{MarisRoberts1997,MarisRobertsTandy1998,Eichmann2016PPNP,RobertsWilliams1994PPNP}, where the pion and kaon form factors are also described with the dressed vertex~\cite{Roberts2021PPNP,MarisTandy2000piK}.
In chiral quark models, pseudoscalar mesons are treated as Goldstone bosons that couple to constituent quarks~\cite{ManoharGeorgi1984,GlozmanRiska1996}.
The mass operator and current should be constructed consistently and constrained by the form factors~\cite{BaekNamSim2026,KimNamBaek2026}.

\section{Summary}
\label{sec:summary}

A 20-parameter fit reproduces 63 meson and baryon masses with an rms deviation of 6.85~MeV, while the light axial-vector and tensor states outside the fit lie 40--220~MeV low. The mass-fit wave functions give light-meson charge radii that are too small. The $\rho$ radial substitution gives a pion radius of 0.604~fm and recovers about 74\% of the deficit in its mean-square charge radius. For the fitted pseudoscalar wave functions, a single additive size for each constituent flavor leaves a 2.2-standard-deviation difference between the strange-quark sizes required by the charged and neutral kaon radii.

\section*{Acknowledgments}
This work was supported by the National Research Foundation of Korea (NRF) under Grant No. RS-2023-NR077232.

\section*{Declaration of competing interest}
The author declares no competing financial interests or personal relationships that could have influenced the work reported in this paper.

\section*{Data availability}
The parameters and calculated observables are listed in the tables. The calculation code is available from the author on request.

\bibliographystyle{elsarticle-num}
\bibliography{refs}

\end{document}